\documentclass[ios]{emulateapj}
\usepackage{color}
\slugcomment{draft \today}
 
\shorttitle{}
\shortauthors{}
\begin{document}
\title{A new density variance - Mach number relation for subsonic and supersonic, isothermal turbulence}
\shorttitle{A new density variance - Mach number relation for turbulent media}
\author{Konstandin, L.\altaffilmark{1,2}, Girichidis, P.\altaffilmark{1,2,3,4}, Federrath, C.\altaffilmark{1,5}, Klessen, R. S.\altaffilmark{1}}
 
\email{email: konstandin@stud.uni-heidelberg.de}
 
\altaffiltext{1}{Zentrum f\"ur Astronomie der Universit\"at Heidelberg,\\Institut f\"ur Theoretische Astrophysik, Albert-Ueberle-Str.2, D-69120 Heidelberg, Germany\\}
\altaffiltext{2}{Member of the International Max Planck Research School for Astronomy and Cosmic Physics at the University of Heidelberg (IMPRS-HD) and the Heidelberg
 Graduate School of Fundamental Physics (HGSFP).\\}
\altaffiltext{3}{Max Planck Institut f\"ur Astrophysik, Karl-Schwarzschild-Str.~1, 85741 Garching, Germany\\}
\altaffiltext{4}{Hamburger Sternwarte, Gojenbergsweg 112, 21029 Hamburg, Germany\\}
\altaffiltext{5}{Monash Centre for Astrophysics (MoCA), School of Mathematical Sciences, Monash University, Vic 3800, Australia\\}

\begin{abstract}
The probability density function (PDF) of the gas density in subsonic and supersonic, isothermal, driven turbulence is analysed with
 a systematic set of hydrodynamical grid simulations with resolutions up to $1024^3$ cells.
We performed a series of numerical experiments with root mean square (r.m.s.) Mach number $\mathcal{M}$ ranging from the nearly
 incompressible, subsonic ($\mathcal{M}=0.1$) to the highly compressible,
 supersonic ($\mathcal{M}=15$) regime.
We study the influence of two extreme cases for the driving mechanism by applying a purely solenoidal (divergence-free) and a purely compressive (curl-free) forcing field
 to drive the turbulence.
We find that our measurements fit the linear relation between the r.m.s.~Mach number and the standard deviation of the density distribution in a wide
 range of Mach numbers, where the proportionality constant depends on the type of the forcing. In addition, we propose a new linear relation between the standard deviation
 of the density distribution $\sigma_{\rho}$ and the standard deviation of the velocity in compressible modes,
 i.e.~the compressible component of the r.m.s.~Mach number $\mathcal{M}_{\mathrm{comp}}$.
In this relation the influence of the forcing is significantly reduced, suggesting a linear relation between $\sigma_{\rho}$ and $\mathcal{M}_{\mathrm{comp}}$, independent of the forcing,
 ranging from the subsonic to the supersonic regime.
\end{abstract}
\section{Introduction}
Understanding the intricate interplay between interstellar turbulence and self-gravity is one of the key problems in star formation theory.
 The supersonic turbulent velocity field is likely responsible for the complex and filamentary density structures observed in molecular clouds.
 It creates dense regions that can become gravitationally unstable and collapse into dense cores and eventually turn into new stars \citep{Elmegreen2004, MacLow2004, McKee2007}.
 Statistical quantities, describing this process, such as the initial mass function, the core mass function \citep{Padoan2002, Hennebelle2008, Hennebelle2009}, and the star formation rate
 \citep{Hennebelle2011, Padoan2011} depend on the standard deviation (std.~dev.) of the density of the molecular cloud. The pioneering works by
 \citet{Padoan1997} and \citet{Passot1998} have shown that the std.~dev.~$\sigma_{\rho}$ of the probability density function(PDF)
 of the mass-density grows proportional to the root mean square (r.m.s.) Mach number $\mathcal{M}$
 of the turbulent flow
\begin{equation}
\sigma_{\rho}/\! \left<\rho \right>_V = b \,\mathcal{M}\,,
\label{eq:Passot1998} 
\end{equation}
where $\left< \rho \right>_V$ is the volume-weighted mean density and $b$ is a proportionality constant.
 \citet{Federrath2008, Federrath2010} explained the dependence of $\sigma_{\rho}$ on $b$ by taking the modes of the forcing into account that drive the turbulent velocity field.
 This model predicts for purely solenoidal forcing $b=1/3$ and for purely compressive forcing $b=1$, and explains
 the large deviations of $b$ ranging from $b=0.26$ to $b=1.05$ in previous works \citep[e.g.][]{Padoan1997, Passot1998, Li2003, Kritsuk2007, Beetz2008, Schmidt2009, Price2011, Konstandin2012,
 Molina2012}.
 We follow up on this work and discuss the physical origin of this dependency and introduce a new relation, similar to equation (\ref{eq:Passot1998}), however, correlating the
 compressible component of the r.m.s~Mach number $\mathcal{M}_{\mathrm{comp}}$ with $\sigma_{\rho}$.\\
In section \ref{sec:Simulation} we explain our numerical setup. We analyse the influence of measuring mass-weighted and volume-weighted distributions in section \ref{subsec:VW_vs_MW},
 the influence of the resolution on our measurements in section \ref{subsec:Resolution} and
 the PDFs of the mass density and the compressible part of the velocity field in section
 \ref{subsec:The_PDF_of_comressive_modes}. In section \ref{subsec:Relation}, we present the linear relations between
 the std.~dev.~of the mass density and the r.m.s.~Mach number. In section \ref{subsec:Physical_origin} we discuss the new relation between the std.~dev.~of the mass density and
 the std.~dev.~of the compressible part of the velocity field. A summary of our results and conclusions is given in section \ref{sec:Summary}.
\section{Simulations and methods}
\label{sec:Simulation}
We use the piecewise parabolic method \citep{Coella1984} implemented in the grid code FLASH3 \citep{Fryxell2000, Dubey2008} to solve the hydrodynamical equations on a uniform three-dimensional grid.
 These equations are the continuity equation
\begin{equation}
 \frac{\partial \rho}{\partial t} +(\textbf{v} \cdot \nabla)\rho=-\rho \nabla \cdot \textbf{v} \,,
\label{eq:continuity}
\end{equation}
the Euler equation with a stochastic forcing term $\mathbf{F}$ per unit mass
\begin{equation}
\frac{\partial \textbf{v}}{\partial t} +(\textbf{v} \cdot \nabla)\textbf{v}=-\frac{ \nabla p}{\rho} + \textbf{F} \, ,
\end{equation}
and the equation of state
\begin{equation}
 p = \kappa \rho^{\Gamma}\,,
\end{equation}
where $\mathbf{v}$ is the velocity field, $s=\ln(\rho/\left< \rho \right>_V)$ is the natural logarithm of the mass density $\rho$, $c_\mathrm{s}$ is the sound speed,
 $p$ the pressure, $\Gamma$ the adiabatic index.
 Since isothermal gas is assumed throughout this study, $\Gamma=1$, the pressure, $p = \rho c_\mathrm{s}^2$, is proportional
 to the mass density with a fixed sound speed $c_\mathrm{s}$.
 These simulations are scale free, so we set
 $\langle \rho \rangle_V = 1$, $c_\mathrm{s}=1$, and the box size of the simulation $L=1$. The numerical simulations are evolved for ten dynamical time scales
 $T = L/\left(2\mathcal{M}c_\mathrm{s}\right)$, where $\mathcal{M}=v_{r.m.s.}/c_\mathrm{s}$ is the r.m.s.~Mach number of the simulations with the r.m.s.~velocity $v_{r.m.s.}$.
 All relevant quantities are stored in intervals of $0.1T$.
 The stochastic forcing field $\mathbf{F}$ has an autocorrelation time equal to the dynamical time scale on the injection scale,
 and varies smoothly in space and time. The forcing field is constructed in Fourier space such that the kinetic energy is injected on the largest scales, where $1< k L / 2\pi< 3$.
 To analyse the influence of different
 modes of the forcing field, we use projection tensors in Fourier space to get a purely divergence-free, ${\nabla} \cdot \textbf{F} = 0$, solenoidal or a purely curl-free,
 $ {\nabla} \times \textbf{F} = 0 $, compressive vector field for the forcing. We adjust the amplitude of the forcing, such that we have
 r.m.s.~Mach numbers $\mathcal{M} = 0.1,\, 0.5,\, 2,\, 5.5,\, 15$ for both types of forcing in the stationary state of fully
 developed turbulence.
 To investigate the effects of numerical viscosity, we study simulations at different resolutions, $128^3$, $256^3$, $512^3$ and $1024^3$.
 The parameters of these simulations are described in \citet{Konstandin2012},
 and a detailed description of the forcing is presented in \citet{Schmidt2009} and \citet{Federrath2010}.
\section{Results}
\label{sec:results}
\begin{figure}[TtHh]
\includegraphics[width=1 \linewidth]{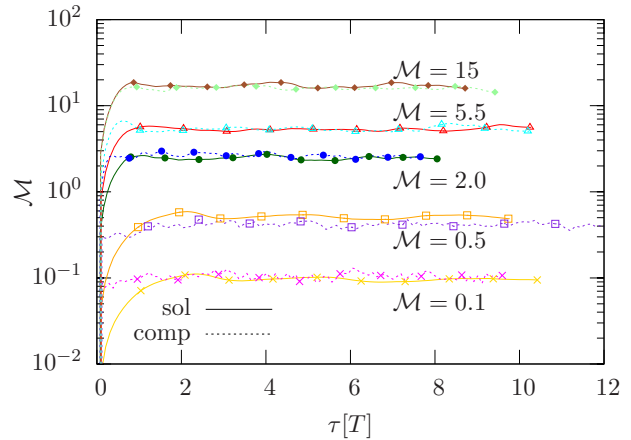}
\caption{R.m.s.~Mach number of all simulations as a function of the dynamical time scale, calculated by averaging over all grid cells for both types of forcing.}
\label{fig:t_RMS_Dens}
\end{figure}

Figure \ref{fig:t_RMS_Dens} shows the time evolution of the r.m.s.~Mach numbers $\mathcal{M}$ in all simulations. The fluid reaches the equilibrium state of fully developed
 turbulence after about two turbulent crossing times $t\approx 2\,T$. We thus average all the following analyses for $2 \leq t/T $.

\subsection{Volume-weighted and mass-weighted probability density functions}
\label{subsec:VW_vs_MW}
It is well-known that the PDF of the logarithm of the mass density $p(s)$ in a turbulent, isothermal medium is close to a Gaussian distribution
 \citep[see e.g.][]{Vazquez1994, Passot1994, Padoan1997, Klessen2000, Kritsuk2007, Federrath2008, Konstandin2012}
\begin{equation}
p(s)=\frac{1}{\sqrt{2\pi}\sigma_{s}}\exp{\left(\frac{-(s-\langle s \rangle)^2}{2\sigma_s^2}\right)}\,.
\end{equation}
\citet{Li2003} showed with the assumption of a Gaussian, volume-weighted PDF of $s$ that the mass-weighted PDF of $s$ is also Gaussian with the same std.~dev.~and with
 a shifted mean value
\begin{equation}
\left<s\right>_V = -\left<s\right>_M = -\frac{{\sigma_s}^2}{2}\,.
\label{eq:s_v_s_m}
\end{equation}
\begin{figure}[TtHh]
\includegraphics[width=1 \linewidth]{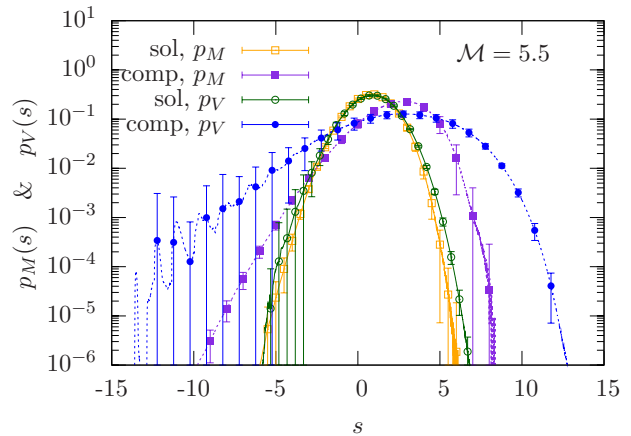}
\caption{Mass-weighted and volume-weighted PDFs of the logarithm of the mass density in the simulations with $\mathcal{M}=5.5$, $1024^3$ grid cells and both types of forcing.}
\label{fig:Grid_PDFs_compaire}
\end{figure}
Figure \ref{fig:Grid_PDFs_compaire} shows the volume- and
 mass-weighted PDFs (the volume-weighted PDF is shifted with $\left<s\right>_M-\left<s\right>_V=\sigma_{s}^2$ for a better comparison) for the simulation with $\mathcal{M} =5.5$
 for both types of forcing. The PDFs are averaged over $81$ time snapshots in the state of fully developed stationary turbulence for $t \geqslant 2T$ and the
 error bars indicate the std.~dev.~of the temporal fluctuations. The variance of the volume-weighted PDFs is
 larger than the variance of the mass-weighted distributions. This effect is stronger for the compressive forcing than for the solenoidal forcing.
 The volume-weighted PDFs show a larger variation with time in the low-density wing of the distribution than the mass-weighted distributions.
 This low-density wing also shows higher probabilities than one would expect from the underlying Gaussian distribution extrapolated from the high density wing.
 This effect is stronger for the compressive than for the solenoidal forcing. We assume that this behaviour is caused by our forcing scheme. 
 As the time correlation of the forcing field is equal to the dynamic time scale on the largest scales, the forcing has enough time to produce very low densities in large regions
 of diverging flows.
 This process causes the volume-weighted PDF of $s$ to have a tail at low densities with higher probabilities than the distribution
 for the case of turbulence, which is not driven on the largest scales.
 With increasing r.m.s.~Mach number and increasing amplitude of the forcing field
 this effect becomes more important and the low-density tail influences the calculated std.~dev.~of these distributions. This effect is less pronounced measuring the mass-weighted
 distributions, as the very low density grid cells carry only little mass. We note that there are other potential processes, which
 could lead to non-Gaussian wings in the PDF, such as turbulent intermittence \citep[e.~g.~][]{Klessen2000}.\\
\begin{figure}[TtHh]
\includegraphics[width=1 \linewidth]{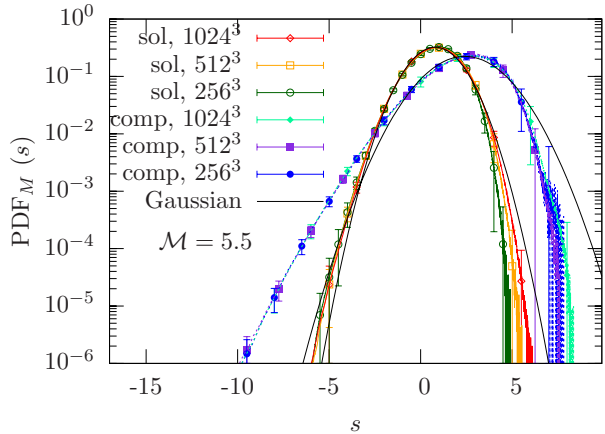}
\caption{Mass-weighted PDFs of $s$ of the simulations for $\mathcal{M}=5.5$, different resolutions and both types of forcing. The black solid lines are Gaussian functions
 with mean value and std.~dev.~calculated with the highest resolution.}
\label{fig:Grid_PDFs_resolutions}
\end{figure}
\subsection{Resolution effects on the probability density functions}
\label{subsec:Resolution}
Figure \ref{fig:Grid_PDFs_resolutions} shows the mass-weighted PDF of the quantity $s$ with an r.m.s.~Mach number $\mathcal{M}=5.5$ and different resolutions.
 The PDF of $s$ shows deviations from the Gaussian shape and a dependency on the resolution only in the high-density tails of the distribution.
 We interpret the deviations of our measured PDFs from the Gaussian distribution in the
 supersonic regime for both types of forcing as a sign of numerical dissipation and finite sampling.
 In the highly supersonic regime the medium is dominated by shock fronts and high-density gradients, which require high resolution to converge.
 As we do not  fully resolve them in the $\mathcal{M}=5.5$ case, an additional dissipation occurs. This effect is stronger in the simulations with compressive forcing
 and becomes stronger with increasing r.m.s.~Mach number for both types of forcing (not shown here).
 However, increasing the resolution has no influence on the deviations from the Gaussian distribution in the low-density tail of the mass-weighted PDFs.\\
\begin{figure*}[TtHh]
\includegraphics[width=1 \linewidth]{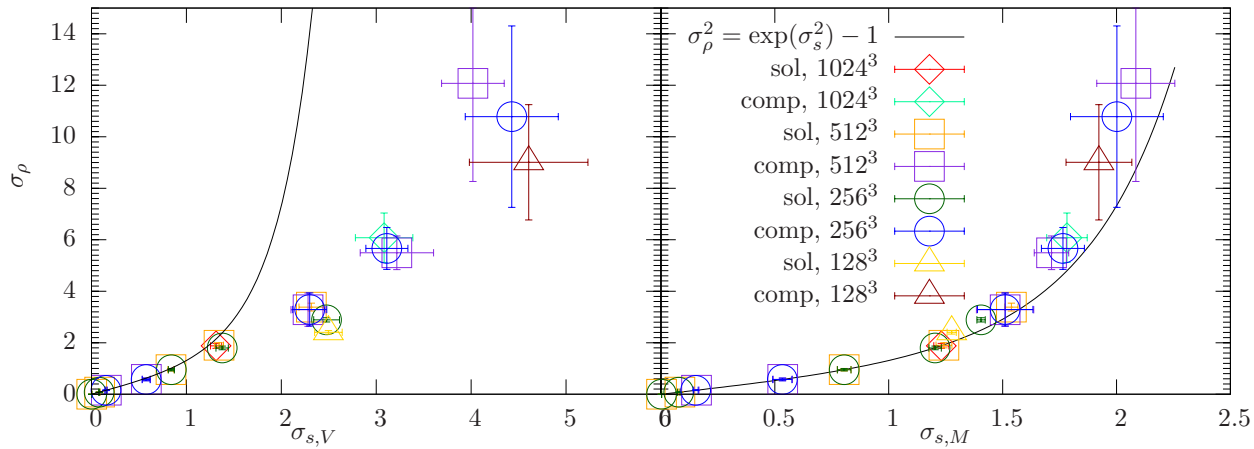}
\caption{Std.~dev.~of the mass-density $\sigma_{\rho}$ as a function of the std.~dev.~of the logarithm of the mass-density $\sigma_{s}$, measured volume-weighted (left panel) and mass-weighted (right panel).
 The deviations of the measurements from the black solid lines, equation (\ref{eq:sig_rho_sig_s}), quantify the deviations
 from a log-normally distributed mass density.} 
\label{fig:sigma_dens_vs_sigma_s}
\end{figure*}
With the assumption of a log-normally distributed mass density, it can be shown that the std.~dev.~of the Gaussian-distributed quantity $s$ is \citep[see][]{Price2011}
\begin{equation}
 \sigma_{s}^2 = \ln{(1+\sigma_{\rho}^2)} \,.
\label{eq:sig_rho_sig_s}
\end{equation}
Figure \ref{fig:sigma_dens_vs_sigma_s} shows $\sigma_{\rho}$ as a function
 of $\sigma_s$ for our volume-weighted (left panel) and mass-weighted (right panel) distributions. The volume- and the mass-weighted measurements of the std.~dev.~of $s$ show increasing
 deviations from equation (\ref{eq:sig_rho_sig_s}) with increasing r.m.s.~Mach numbers for both types of forcing. However, the deviations are smaller in the mass-weighted case than
 in the volume-weighted one. The assumption of Gaussianity, which is implied in equation (\ref{eq:sig_rho_sig_s}), is better fulfilled for the mass-weighted case.
 Figure \ref{fig:sigma_dens_vs_sigma_s} also shows that our measurements with $\mathcal{M}=15$ are not converged with resolution for both types of forcing.
 Our measurements are in agreement with \citet{Price2011}, who showed that direct measurements of $\sigma_{\rho}$ show a stronger dependency on resolution
 than measurements of $\sigma_{s}$.\\
 All volume-weighted measurements show a clear trend towards the relation (\ref{eq:sig_rho_sig_s}) with increasing resolution.
 However, the data points do not fit relation (\ref{eq:sig_rho_sig_s}) for $\mathcal{M}=15$ with solenoidal forcing and in all the supersonic cases with compressive forcing, 
 although the data points with $\mathcal{M}=2$ and $\mathcal{M}=5.5$ with compressive forcing are nearly converged with resolution.
 This can be explained with the non-Gaussian, low-density wing of the distributions, which does not depend on the resolution.
 Considering that the std.~dev.~$\sigma_{s,M}$ of the mass-weighted PDF is more compatible with the scaling for a log-normal PDF, equation (\ref{eq:sig_rho_sig_s}), and that
 the resolution dependence of $\sigma_{s,M}$ is weaker that for $\sigma_{s,V}$, we prefer to use $\sigma_{s,M}$ as estimate for the turbulent density fluctuations in the following.
\subsection{The probability density function of the density and of the compressible modes in the velocity field}
\label{subsec:The_PDF_of_comressive_modes}
\begin{figure*}[TtHh]
\includegraphics[width=1 \linewidth]{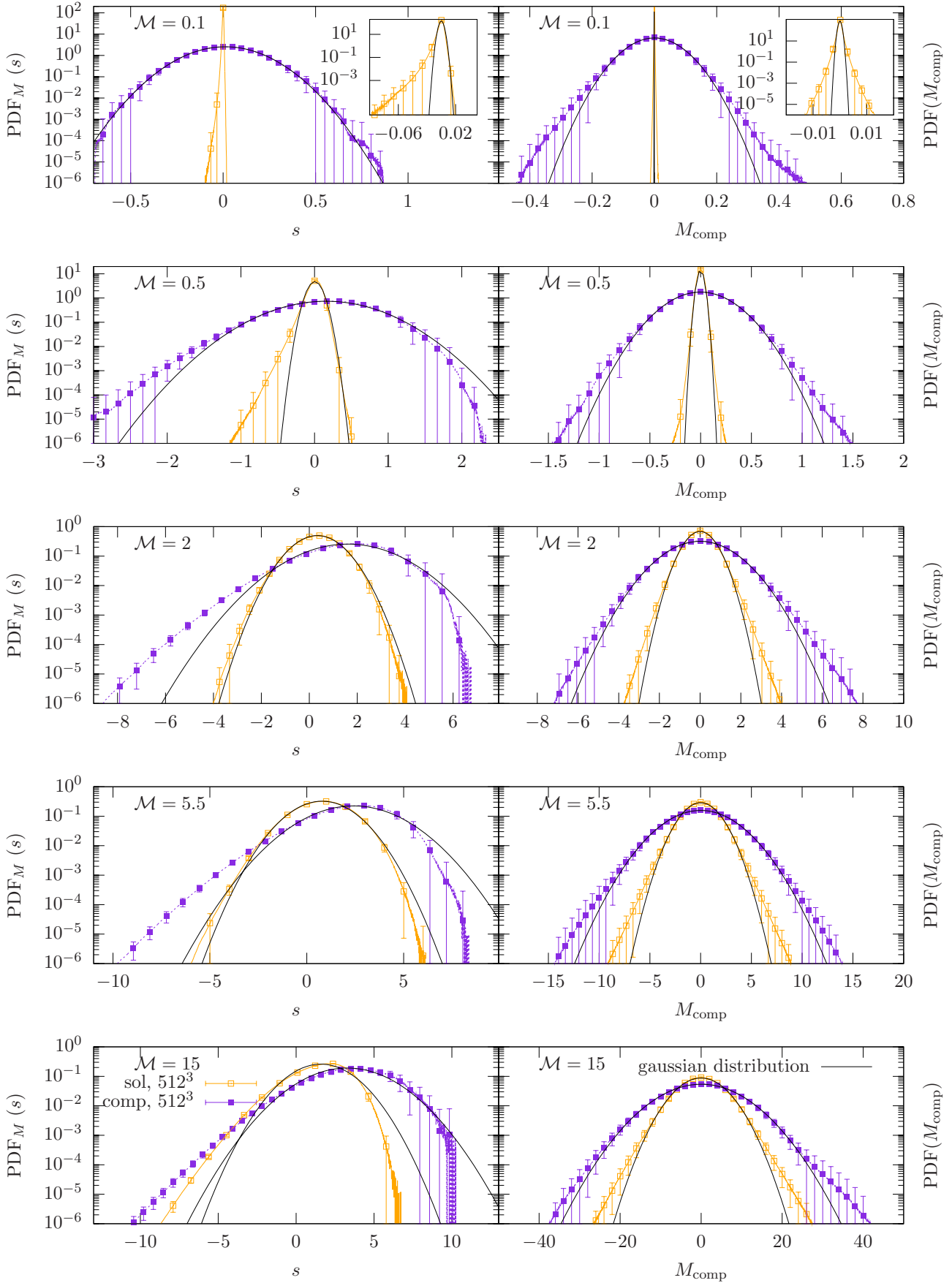}
\caption{The mass-weighted PDFs of the logarithm of the mass density (left panels) and the compressible part of the local Mach number (right panels)
 for different r.m.s.~Mach numbers, resolutions and both types of forcing. In the inset, a magnification of the PDFs obtained with solenoidal forcing for $\mathcal{M}=0.1$ are shown.
 The error bars in each panel indicate the std.~dev.~of the temporal fluctuations.}
\label{fig:Grid_PDFs_density_and_vL}
\end{figure*}

Figure \ref{fig:Grid_PDFs_density_and_vL} shows the mass-weighted PDFs of the quantity $s$ (left panels) and the volume-weighted PDFs of the compressible modes of the
 velocity field normalised to the sound speed $M_{\mathrm{comp}} = v_{\mathrm{comp}} /c_s$ (right panels) for different r.m.s.~Mach numbers and both types of forcing.
 The PDFs of the logarithm of the density largely follow Gaussian distributions for all supersonic r.m.s.~Mach numbers.
 We added Gaussian functions (black solid lines), with the first- and second-order moments calculated from our distributions in figure \ref{fig:Grid_PDFs_density_and_vL}.
 The high-density tails of the distributions show deviations from the Gaussian shape, which increase with increasing r.m.s.~Mach number.
 Also the deviations from the Gaussian distribution in the low-density tail, as discussed in section \ref{subsec:VW_vs_MW},
 get more pronounced with increasing r.m.s.~Mach number.
 Thereby, we have large deviations of our measurement from the Gaussian distributions in the $\mathcal{M}=15$ case and the calculated std.~dev.~does not
 correspond to the std.~dev.~of the underlying Gaussian distribution.\\
 The density distributions of the simulations driven by solenoidal forcing in the subsonic regime
 show significant deviations from the log-normal shape, which become stronger as $\mathcal{M}$ decreases.
 These distributions become more asymmetric and more peaked. The different behaviour of the PDFs
 in the subsonic regime especially for the solenoidal forcing is caused by the different physical processes acting here. In the subsonic regime sound
 waves transfer information faster
 than the averaged flow of the medium, such that the thermal pressure increases before two converging flows
 can collide. This process prevents colliding flows 
 from producing high-density regions and causes the sharp edge at the high-density wing of the distributions. The thermal pressure also
 decelerates the velocities in compressible modes, such that the PDF of $M_{\mathrm{comp}}$ also shows a narrow, peaky and intermittent behaviour for the solenoidal forcing.
 This process is just visible for solenoidal forcing, because in the compressive forcing case the velocities in
 compressible modes are  re-injected by the forcing to hold the r.m.s.~Mach number constant.
 This is the reason why the thermal pressure does not have such a strong influence there.\\
The right panels of figure \ref{fig:Grid_PDFs_density_and_vL} show the PDFs of $M_{\mathrm{comp}}$, where $M_{\mathrm{comp}}$ is calculated in Fourier space
 and averaged over the three directions of the coordinate system $x, y, z$ in real space. The distributions of $M_{\mathrm{comp}}$ are symmetric with zero mean and have an
 increasing std.~dev.~with increasing r.m.s.~Mach number. The distributions obtained with compressive forcing are always broader than with solenoidal forcing at the same r.m.s.~Mach number.
 The PDFs of $M_{\mathrm{comp}}$ are Gaussian (black solid lines) with deviations in both wings. These are the signpost of turbulent intermittency.
 The deviations do not show a clear trend with the r.m.s.~Mach number.\\
 The PDF of $M_{\mathrm{comp}}$ obtained with solenoidal forcing in the subsonic regime with $\mathcal{M}=0.1$
 shows the strongest deviations from the Gaussian shape with a narrow, peaky, intermittent
 distribution. These deviations are caused by the thermal pressure, as discussed above.\\

\subsection{Relation between the r.m.s.~Mach number and the standard deviation of the density}
\label{subsec:Relation}
In \citet{Padoan1997} and \citet{Passot1998} the authors found that the std.~dev.~of the PDF of the mass density $\sigma_{\rho}$
 is proportional to the r.m.s.~Mach number in a turbulent flow.
The std.~dev.~of the mass density is an important quantity especially in astrophysics, where the Mach number dependency of density fluctuations is used to derive analytic expressions for
 the core mass function (CMF) and
 the stellar initial mass function (IMF) \citep[e.g.,~][]{Padoan2002, Hennebelle2008, Hennebelle2009}. On galactic scales it
 is used to reproduce the Kennicutt-Schmidt relation \citep{Tassis2007},
 and \citet{Elmegreen2008} suggests that the star formation efficiency is a function of the density PDF.
\begin{figure*}[TtHh]
\includegraphics[width=1 \linewidth]{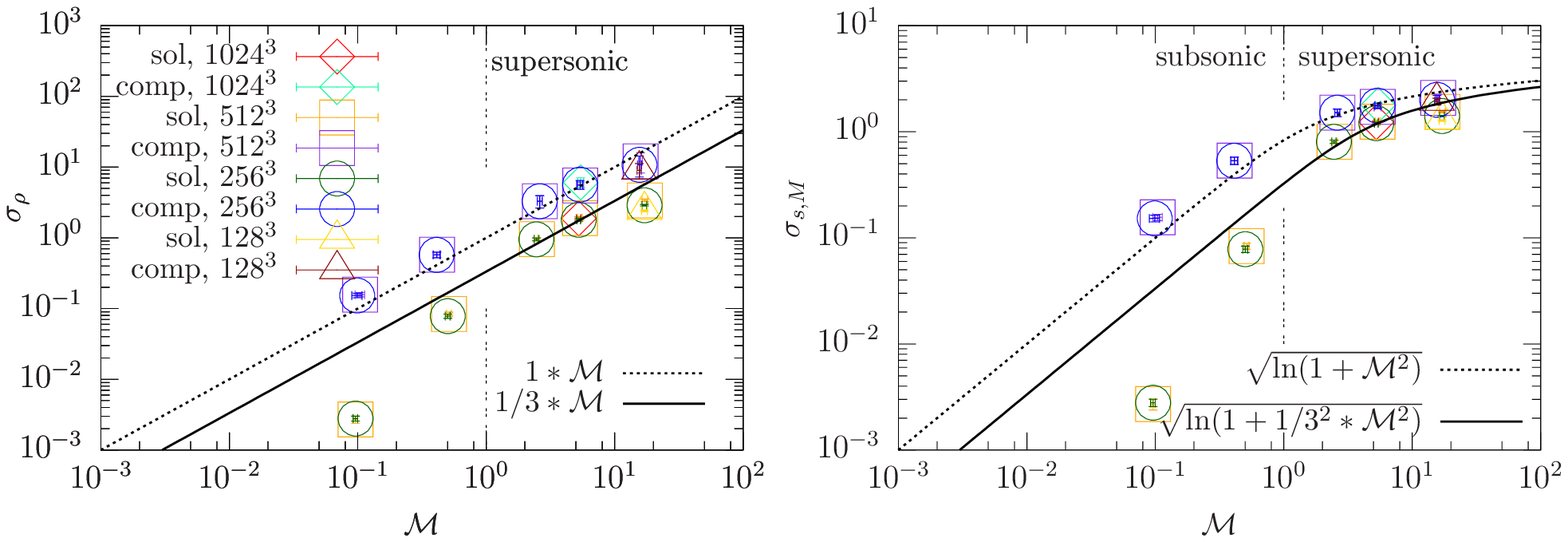}\\
\includegraphics[width=1 \linewidth]{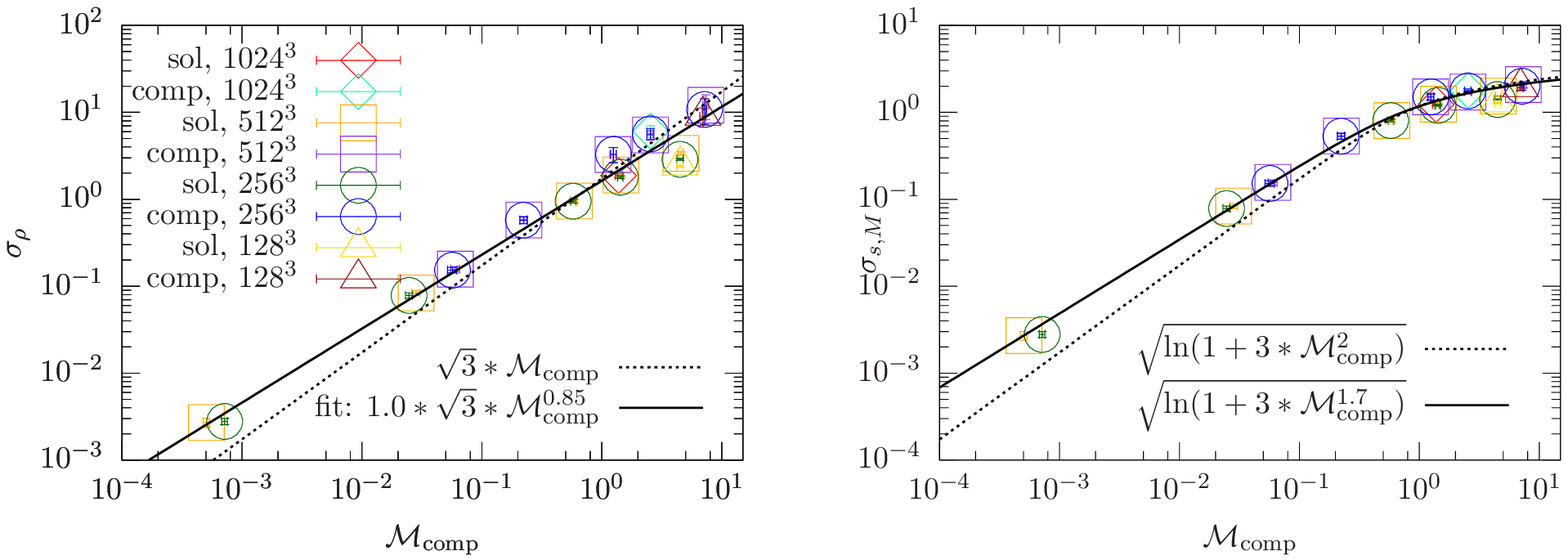}
\caption{The std.~dev.~of the distribution of the mass density (left) and the std.~dev.~of the distribution of the logarithm of the mass density (right) as a function of the r.m.s.~Mach number (upper panels)
 and as a function of the compressible part of the r.m.s.~Mach number (lower panels). In the upper panels the lines correspond to the model of \citet{Federrath2010} with $b=1/3$
 for solenoidal forcing and $b=1$ for compressive forcing. In the lower panels the solid lines correspond to a two-parameter fit and the dotted line corresponds to a linear relation between
 the std.~dev.~of the mass density and the std.~dev.~of the compressible part of the r.m.s.~Mach number with a proportionality constant $\sqrt{3}$.}
\label{fig:result}
\end{figure*}
Figure \ref{fig:result} (upper left panel) shows the measured std.~dev.~of the mass density as a function of the r.m.s.~Mach number for different resolutions and both types of forcing.
 The error bars in each panel of figure \ref{fig:result} indicate the std.~dev.~of the temporal fluctuations of the measured quantities. They do not include any potential systematic
 errors stemming from, e.g., the numerical scheme or implementation of the forcing algorithm.
 Thus, we interpret the error bars as a lower limit of the real uncertainty.
 The dotted and dashed-dotted lines correspond to the model of \citet{Federrath2010}, which describes the proportionality parameter $b$ as a function of the turbulent forcing.
 This model predicts for solenoidal forcing $b=1/3$ and for compressive forcing $b=1$.
 Our measurements agree with the model of \citet{Federrath2010} in the supersonic case for both types of forcing. We see small deviations from the model in the
 simulations with $\mathcal{M}=15$, which is caused by our limited resolution (see figure \ref{fig:sigma_dens_vs_sigma_s}). 
 The std.~dev.s of the density distribution of the simulation with solenoidal forcing
 are smaller than the prediction of the model in the subsonic case. In
 the subsonic regime, the deviations are caused by the thermal pressure, which damps
 density variations and compressible modes of the velocity field and reduces the measured std.~dev.~below the model prediction as discussed in section \ref{subsec:The_PDF_of_comressive_modes}.
 The upper right panel of figure \ref{fig:result} shows the mass-weighted, logarithmic std.~dev.~$\sigma_{s,M}$ as a function of the r.m.s.~Mach number.
 The dotted and dashed-dotted lines correspond to the standard model for the logarithmic density variance,
\begin{equation} \label{eq:sigs}
\sigma_s^2 = \ln(1+b^2\mathcal{M}^2)
\end{equation}
with $b=1/3$ for solenoidal and $b=1$ for compressive forcing. Equation~(\ref{eq:sigs}) follows from equations
 (\ref{eq:sig_rho_sig_s}) and (\ref{eq:Passot1998}) and was recently derived analytically by \citet{Molina2012}
 using the shock-jump conditions and averaging over an ensemble of shock waves.
 The deviations of our numerical data from this standard model are only significant for solenoidal
 forcing in the subsonic regime, while our data are in excellent agreement with equation~(\ref{eq:sigs})
 for both solenoidal and compressive forcing in the supersonic regime, given our resolution dependence of the $\mathcal{M}=15$ data points. 
%
 Our results are in agreement with \citet{Kowal2007}, who found deviations from the linear relation with $\sigma_{\rho}$ in the subsonic regime with solenoidal forcing,
 and with \citet{Passot1998}, who analysed one-dimensional
 simulations with only compressive forcing and $0.5 \leq \mathcal{M} \leq 3$ and found the linear relation between $\mathcal{M}$ and $\sigma_{\rho}$ with $b=1$.
 \citet{Price2011} analysed three dimensional simulations with purely solenoidal forcing and r.m.s.~Mach numbers in between $2 \leq \mathcal{M} \leq 20$
 and found $b=1/3$ in excellent agreement with our result.
 As they did not analyse the subsonic regime with solenoidal forcing they did not observe the large deviations in the subsonic regime.
 Our analysis complements these studies with measurements in both the subsonic and supersonic regime and for purely compressive forcing. 
\subsection{Physical origin of density fluctuations in turbulent flows}
\label{subsec:Physical_origin}
 Studying the continuity equation (\ref{eq:continuity}), one can argue that variations of the density can only be caused by the divergence of the velocity field.
 Given that a vector field can be decomposed in a gradient field and a rotation field and that the divergence of a rotation field vanishes, we conclude that the density variations
 can only be caused by the compressible modes of the velocity. For this reason 
 we replaced the r.m.s.~Mach number, which is
 in fact the std.~dev.~of the velocity distribution, with the compressible part of the r.m.s.~Mach number, $\mathcal{M}_{\mathrm{comp}}$,
 in the lower panels of figure \ref{fig:result}.
 The data points show a clear correlation. The different behaviour of the simulations driven with solenoidal and compressive
 forcing are significantly reduced.
 We added in figure~\ref{fig:result} a function (dotted line) for the relation $\sigma_{\rho} = \sqrt{3}\mathcal{M}_{\mathrm{comp}}$, which is the simplest model for this relation
 assuming isotropy.
 The factor of $\sqrt{3}$ is due to the fact that we use the distribution of the compressible modes of the velocity field averaged over the three directions of the coordinate system
\begin{equation}
\mathcal{M}_{\mathrm{comp}}^{\mathrm{tot}} = \sqrt{\mathcal{M}_{\mathrm{comp},\,x}^2+ \mathcal{M}_{\mathrm{comp},\,y}^2+ \mathcal{M}_{\mathrm{comp},\,z}^2}
\end{equation}
\[
  =\sqrt{3}  \mathcal{M}_{\mathrm{comp}}\,.
\]
A similar model has also been suggested by \citet{Federrath2010},
 where the parameter $b$ in equations~(\ref{eq:Passot1998}) and~(\ref{eq:sigs}) was approximated by $\sqrt{D}\left<\Psi\right>$
 with the spatial dimension $D$ of the turbulence ($D=3$ in our case) and the so-called compressive ratio $\left<\Psi\right>$,
 which is the ratio of compressible to total velocity fluctuations.
 This and our simple model fits the data, but shows deviations for the simulations with solenoidal forcing and the lowest and highest Mach numbers. The deviations
 for the $\mathcal{M}=15$ simulation are again caused by the resolution dependency of $\sigma_{\rho}$. 
 Additionally, we perform a fit of our data (black solid line) with two free parameters, $\sigma_{\rho}=\alpha \sqrt{3}\mathcal{M}_{\mathrm{comp}}^{\beta}$ for the density relation. 
 We obtain a normalisation $\alpha=1.0 \pm 0.1$ and a slope $\beta=0.85 \pm 0.04$. For the $s$-relation we transform the fitted function with equation (\ref{eq:sig_rho_sig_s}).
 The measurements of the std.~dev.~of the density have larger deviations from the model as the measurements of the std.~dev.~of $s$.
 However, the model fits the measurements in both cases and provides a good description for
 the data points in the subsonic regime with solenoidal forcing, which are strongly influenced by sound waves.
 We conclude that the thermal pressure damps the velocities in compressible modes in a way that the relation between the velocities in compressible modes
 and the density variations in a turbulent medium is in a statistical equilibrium state, even if the medium is strongly influenced by sound waves.
 The deviation of the scaling exponent from the simple model can be interpreted as additional dissipative effects, which are proportional to $\mathcal{M}_{\mathrm{comp}}$. 
 An example for these physical processes are shocks, which cause deviations from
 the log-normal distribution of the density PDF.
 However, systematic errors with a dependency on the r.m.s.~Mach number could also cause deviations from the linear scaling and would
 be another possible interpretation for our fitted scaling exponent.\\
 The shown relation between the std.~dev.~of the density and the compressible part of the r.m.s.~Mach number in principle enables us to measure the
 kinetic energy in compressible modes in giant molecular clouds, without knowing the absolute r.m.s.~Mach number, the driving mechanism or the sound speed.
 The relations shown in the bottom panels of figure \ref{fig:result} are valid in both, the subsonic and supersonic regime.

\section{Summary and conclusions}
\label{sec:Summary}
We have investigated the influence of solenoidal (divergence-free) and compressive (curl-free) forcing on the PDF of the mass density in subsonic and supersonic turbulence
 with a set of three-dimensional numerical simulations. We analysed the relation between the std.~dev.~of the mass density distribution and the r.m.s.~Mach number.
 We found a new relation between the std.~dev.~of the mass density and the std.~dev.~of the compressible part of the velocity field.
 Our main results are as follows:
\begin{itemize}
 \item Compressive forcing yields mass density PDFs with std.~dev.s proportional to
 the r.m.s.~Mach number with $b=1$. For solenoidal forcing, we measure $b=1/3$ in the supersonic regime.
 Our findings are in agreement with previous studies, which however only explored different subsets of the full parameter space investigated here.
 We also found deviations of our measurements from the linear relation with solenoidal forcing in the
 subsonic regime. These deviations from the linear relation can be explained with sound waves,
 which damp the faint compressible velocities and prevent the medium from producing over-densities. 
 \item We found a unique relation between the std.~dev.~of the mass density and the compressible modes of
 the velocity field with a fit to our data. Our new relation is independent of the driving
 mechanism and still holds in the subsonic regime, where the flow is mainly influenced by sound waves.
 It does not show a strong influence on the resolution and other effects, which may cause a non-Gaussian distribution of the density.
 \item Our relation enables us for the first time to measure the kinetic energy in compressible modes in units of
 the sound speed, without knowing the r.m.s.~Mach number, the driving mechanism or the sound speed of the medium.
 This measurement can be used to distinguish between subsonic and supersonic compressive turbulent motions. 
 It will in principle allow us to measure the composition of the kinetic energy in the interstellar medium by combining independent measurements of the total
 r.m.s.~Mach number \citep[e.g.][]{Burkhart2009} and the std.~dev.~of the density distribution \citep[][]{Brunt2010a, Brunt2010b, Schneider2012}. 
\end{itemize}
\begin{acknowledgments}      
 L.K.~andP.G.~acknowledge financial support by the International Max Planck Research School for Astronomy and Cosmic Physics (IMPRS-A) and the Heidelberg Graduate School of Fundamental
 Physics (HGSFP). The HGSFP is funded by the Excellence Initiative of the German Research Foundation DFG GSC 129/1.
 L.K., P.G., C.F., and
 R.S.K. acknowledge subsidies from the Baden-W\"{u}rttemberg-Stiftung via the program {\em Internationale Spitzenforschung II} (grant P-LS-SPII/18)
 from the German Bundesministerium f\"ur Bildung und Forschung via the ASTRONET project STAR FORMAT (grant 05A09VHA).
 P.G. acknowledges the support by the Max-Planck Institut f\"{u}r Astrophysik in Garching.
 C.F.~acknowledges funding provided by the Australian Research Council under the Discovery Projects scheme (grant no.~DP110102191).
 R.S.K. furthermore gives thanks for subsidies from the Deutsche Forschungsgemeinschaft (DFG) under grants KL 1358/10, and KL 1358/11
 and via the SFB 881 'The Milky Way System',
 as well as from a Frontier grant of Heidelberg University sponsored by the German Excellence Initiative.
 Supercomputing time at the Leibniz Rechenzentrum
 (grant no.~h0972 and pr32lo), and at the Forschungszentrum J\"ulich (grant no.~hhd20) are gratefully acknowledged.
 The software used in this work was in part developed by the DOE-supported ASC / Alliance Center for Astrophysical Thermonuclear Flashes at the University of Chicago.\\

\end{acknowledgments}
\bibliographystyle{apj}
\bibliography{citations.bib}

\begin{thebibliography}{32}
\expandafter\ifx\csname natexlab\endcsname\relax\def\natexlab#1{#1}\fi

\bibitem[{{Beetz} {et~al.}(2008){Beetz}, {Gollwitzer}, {Richter}, \&
  {Rehberg}}]{Beetz2008}
{Beetz}, A., {Gollwitzer}, C., {Richter}, R., \& {Rehberg}, I. 2008, Journal of
  Physics Condensed Matter, 20, 204109

\bibitem[{{Brunt}(2010)}]{Brunt2010b}
{Brunt}, C.~M. 2010, Astronomy and Astrophysics, 513, A67

\bibitem[{{Brunt} {et~al.}(2010){Brunt}, {Federrath}, \& {Price}}]{Brunt2010a}
{Brunt}, C.~M., {Federrath}, C., \& {Price}, D.~J. 2010, Monthly Notices of the
  Royal Astronomical Society, 403, 1507

\bibitem[{{Burkhart} {et~al.}(2009){Burkhart}, {Falceta-Gon{\c c}alves},
  {Kowal}, \& {Lazarian}}]{Burkhart2009}
{Burkhart}, B., {Falceta-Gon{\c c}alves}, D., {Kowal}, G., \& {Lazarian}, A.
  2009, The Astrophysical Journal, 693, 250

\bibitem[{{Colella} \& {Woodward}(1984)}]{Coella1984}
{Colella}, P., \& {Woodward}, P.~R. 1984, Journal of Computational Physics, 54,
  174

\bibitem[{{Dubey} {et~al.}(2008){Dubey}, {Fisher}, {Graziani}, {Jordan},
  {Lamb}, {Reid}, {Rich}, {Sheeler}, {Townsley}, \& {Weide}}]{Dubey2008}
{Dubey}, A., {Fisher}, R., {Graziani}, C., {Jordan}, IV, G.~C., {Lamb}, D.~Q.,
  {Reid}, L.~B., {Rich}, P., {Sheeler}, D., {Townsley}, D., \& {Weide}, K.
  2008, in Astronomical Society of the Pacific Conference Series, Vol. 385,
  Numerical Modeling of Space Plasma Flows, ed. N.~V. {Pogorelov}, E.~{Audit},
  \& G.~P. {Zank}, 145

\bibitem[{{Elmegreen}(2008)}]{Elmegreen2008}
{Elmegreen}, B.~G. 2008, The Astrophysical Journal, 672, 1006

\bibitem[{{Elmegreen} \& {Scalo}(2004)}]{Elmegreen2004}
{Elmegreen}, B.~G., \& {Scalo}, J. 2004, Annual Review of Astronomy and
  Astrophysics, 42, 211

\bibitem[{{Federrath} {et~al.}(2008){Federrath}, {Klessen}, \&
  {Schmidt}}]{Federrath2008}
{Federrath}, C., {Klessen}, R.~S., \& {Schmidt}, W. 2008, The Astrophysical
  Journal, Letters, 688, L79

\bibitem[{{Federrath} {et~al.}(2010){Federrath}, {Roman-Duval}, {Klessen},
  {Schmidt}, \& {Mac Low}}]{Federrath2010}
{Federrath}, C., {Roman-Duval}, J., {Klessen}, R.~S., {Schmidt}, W., \& {Mac
  Low}, M. 2010, Astronomy and Astrophysics, 512, A81

\bibitem[{{Fryxell} {et~al.}(2000){Fryxell}, {Olson}, {Ricker}, {Timmes},
  {Zingale}, {Lamb}, {MacNeice}, {Rosner}, {Truran}, \& {Tufo}}]{Fryxell2000}
{Fryxell}, B., {Olson}, K., {Ricker}, P., {Timmes}, F.~X., {Zingale}, M.,
  {Lamb}, D.~Q., {MacNeice}, P., {Rosner}, R., {Truran}, J.~W., \& {Tufo}, H.
  2000, Astronomy and Astrophysics, Supplement, 131, 273

\bibitem[{{Hennebelle} \& {Chabrier}(2008)}]{Hennebelle2008}
{Hennebelle}, P., \& {Chabrier}, G. 2008, The Astrophysical Journal, 684, 395

\bibitem[{{Hennebelle} \& {Chabrier}(2009)}]{Hennebelle2009}
---. 2009, The Astrophysical Journal, 702, 1428

\bibitem[{{Hennebelle} \& {Chabrier}(2011)}]{Hennebelle2011}
---. 2011, The Astrophysical Journal, Letters, 743, L29

\bibitem[{{Klessen}(2000)}]{Klessen2000}
{Klessen}, R.~S. 2000, The Astrophysical Journal, 535, 869

\bibitem[{{Konstandin} {et~al.}(2012){Konstandin}, {Federrath}, {Klessen}, \&
  {Schmidt}}]{Konstandin2012}
{Konstandin}, L., {Federrath}, C., {Klessen}, R.~S., \& {Schmidt}, W. 2012,
  Journal of Fluid Mechanics, 692, 183

\bibitem[{{Kowal} {et~al.}(2007){Kowal}, {Lazarian}, \&
  {Beresnyak}}]{Kowal2007}
{Kowal}, G., {Lazarian}, A., \& {Beresnyak}, A. 2007, The Astrophysical
  Journal, 658, 423

\bibitem[{{Kritsuk} {et~al.}(2007){Kritsuk}, {Norman}, {Padoan}, \&
  {Wagner}}]{Kritsuk2007}
{Kritsuk}, A.~G., {Norman}, M.~L., {Padoan}, P., \& {Wagner}, R. 2007, The
  Astrophysical Journal, 665, 416

\bibitem[{{Li} {et~al.}(2003){Li}, {Klessen}, \& {Mac Low}}]{Li2003}
{Li}, Y., {Klessen}, R.~S., \& {Mac Low}, M. 2003, The Astrophysical Journal,
  592, 975

\bibitem[{{Mac Low} \& {Klessen}(2004)}]{MacLow2004}
{Mac Low}, M.-M., \& {Klessen}, R.~S. 2004, Reviews of Modern Physics, 76, 125

\bibitem[{{McKee} \& {Ostriker}(2007)}]{McKee2007}
{McKee}, C.~F., \& {Ostriker}, E.~C. 2007, Annual Review of Astronomy and
  Astrophysics, 45, 565

\bibitem[{{Molina} {et~al.}(2012){Molina}, {Glover}, {Federrath}, \&
  {Klessen}}]{Molina2012}
{Molina}, F.~Z., {Glover}, S.~C.~O., {Federrath}, C., \& {Klessen}, R.~S. 2012,
  ArXiv e-prints-1203.2117

\bibitem[{{Padoan} \& {Nordlund}(2002)}]{Padoan2002}
{Padoan}, P., \& {Nordlund}, {\AA}. 2002, The Astrophysical Journal, 576, 870

\bibitem[{{Padoan} \& {Nordlund}(2011)}]{Padoan2011}
---. 2011, The Astrophysical Journal, 730, 40

\bibitem[{{Padoan} {et~al.}(1997){Padoan}, {Nordlund}, \& {Jones}}]{Padoan1997}
{Padoan}, P., {Nordlund}, A., \& {Jones}, B.~J.~T. 1997, Monthly Notices of the
  Royal Astronomical Society, 288, 145

\bibitem[{{Passot} \& {V{\'a}zquez-Semadeni}(1998)}]{Passot1998}
{Passot}, T., \& {V{\'a}zquez-Semadeni}, E. 1998, Physical Review E, 58, 4501

\bibitem[{{Passot} {et~al.}(1994){Passot}, {V{\'a}zquez-Semadeni}, \&
  {Pouquet}}]{Passot1994}
{Passot}, T., {V{\'a}zquez-Semadeni}, E.~C., \& {Pouquet}, A. 1994, {A
  turbulent model for the interstellar medium.}, ed. {Franco, J., Lizano, S.,
  Aguilar, L., \& Daltabuit, E.}, 246

\bibitem[{{Price} {et~al.}(2011){Price}, {Federrath}, \& {Brunt}}]{Price2011}
{Price}, D.~J., {Federrath}, C., \& {Brunt}, C.~M. 2011, The Astrophysical
  Journal, Letters, 727, L21

\bibitem[{{Schmidt} {et~al.}(2009){Schmidt}, {Federrath}, {Hupp}, {Kern}, \&
  {Niemeyer}}]{Schmidt2009}
{Schmidt}, W., {Federrath}, C., {Hupp}, M., {Kern}, S., \& {Niemeyer}, J.~C.
  2009, Astronomy and Astrophysics, 494, 127

\bibitem[{{Schneider} {et~al.}(2012){Schneider}, {Csengeri}, {Hennemann},
  {Motte}, {Didelon}, {Federrath}, {Bontemps}, {Di Francesco}, {Arzoumanian},
  {Minier}, {Andr{\'e}}, {Hill}, {Zavagno}, {Nguyen-Luong}, {Attard},
  {Bernard}, {Elia}, {Fallscheer}, {Griffin}, {Kirk}, {Klessen}, {K{\"o}nyves},
  {Martin}, {Men'shchikov}, {Palmeirim}, {Peretto}, {Pestalozzi}, {Russeil},
  {Sadavoy}, {Sousbie}, {Testi}, {Tremblin}, {Ward-Thompson}, \&
  {White}}]{Schneider2012}
{Schneider}, N., {Csengeri}, T., {Hennemann}, M., {Motte}, F., {Didelon}, P.,
  {Federrath}, C., {Bontemps}, S., {Di Francesco}, J., {Arzoumanian}, D.,
  {Minier}, V., {Andr{\'e}}, P., {Hill}, T., {Zavagno}, A., {Nguyen-Luong}, Q.,
  {Attard}, M., {Bernard}, J.-P., {Elia}, D., {Fallscheer}, C., {Griffin}, M.,
  {Kirk}, J., {Klessen}, R., {K{\"o}nyves}, V., {Martin}, P., {Men'shchikov},
  A., {Palmeirim}, P., {Peretto}, N., {Pestalozzi}, M., {Russeil}, D.,
  {Sadavoy}, S., {Sousbie}, T., {Testi}, L., {Tremblin}, P., {Ward-Thompson},
  D., \& {White}, G. 2012, Astronomy and Astrophysics, 540, L11

\bibitem[{{Tassis}(2007)}]{Tassis2007}
{Tassis}, K. 2007, Monthly Notices of the Royal Astronomical Society, 382, 1317

\bibitem[{{Vazquez-Semadeni}(1994)}]{Vazquez1994}
{Vazquez-Semadeni}, E. 1994, The Astrophysical Journal, 423, 681

\end{thebibliography}
\end{document}